\documentstyle[12pt,a4j]{article}
\begin{document}
\setlength{\baselineskip}{2.0pc}
\parindent = 17pt
%
\par
\begin{center}
\begin{large}
\begin{bf}
\vspace*{1.5cm}
Eigenfunctions for SU($\nu$) particles

with $1/r^2$ interaction in harmonic confinement
\end{bf}
\end{large}
\vskip 1.0cm
Karel Vacek and Ayao Okiji \par
\vskip 0.1cm
\begin{it}
Department of Applied Physics, Osaka University, Suita, Osaka 565,
Japan
\par
\end{it}
\vskip 0.1cm
Norio Kawakami \par
\vskip 0.1cm
\begin{it}
Yukawa Institute for Theoretical Physics, Kyoto University, Kyoto 606,
Japan
\par
\end{it}
\vskip 0.3cm
(Received \hskip5cm )
\end{center}
\vskip 0.5cm
\par
%
\centerline {\bf ABSTRACT } \par
\vskip 0.1cm
We find a set of the exact eigenfunctions
which provide the energy spectrum
for the quantum $N$-body Calogero-Sutherland model
for fermions or bosons with SU($\nu$) spin degrees of freedom
moving in a harmonic confinement potential.
The eigenfunctions are explicitly constructed as a simple product of
the Jastrow wave function for the ground state and
the Hermite polynomials introduced to generate the excited states.
The corresponding energy spectrum is given by
a sum of the correlated ground-state energy
and of the excited-state energy for SU($\nu$)
free particles in a harmonic well.

\vskip 0.3cm
%

\newpage

The integrable systems with inverse-square ($1/r^2$) interaction
in one dimension were introduced by Calogero \cite{cal69} and
Sutherland \cite{sut71},
and more recently extended to the spin chain by Haldane \cite{hal88}
and Shastry \cite{sha88}.
Models with periodic boundary conditions
have been then  studied intensively
and the corresponding wavefunctions have been constructed explicitly
\cite{kur91,kaw91,hah92,kaw92,wan92,pol92,fow93,hik93,sut93}.

For the $1/r^2$ systems with harmonic confinement potential
\cite{sut71,pol92,bri92,fra93,min93,vac93},
the formal algebraic structure has been clarified recently
via the construction of appropriate annihilation and creation
operators \cite{pol92,bri92}, which enables one to prove the
integrability
for a class of the SU($\nu$) models with confinement \cite{min93}.
Such an operator formalism is elegant,
giving the answer about the spectrum and the degeneracy of
the energy levels.
%
%
However, we still miss expressions for
the eigenfunctions of excited states for the
$1/r^2$ system with harmonic confinement.
The systematic construction of eigenfunctions
should also provide a microscopic foundation of the
renormalized-harmonic oscillator hypothesis \cite{kaw93b}
which is based on a variant of the asymptotic Bethe ansatz method
\cite{sut71,hal88,kaw92}.

Several authors attempted
to construct explicitly the wave functions for
the excited states of the Calogero-Sutherland model
with confinement, but this problem
has not been solved generally even for the single-component case.
Calogero \cite{cal69} obtained early results for the eigenfunctions of
the system containing $N=3$ and $N=4$ particles.
Calogero's attempt to obtain the solution for the $N$-body
system has left open the problem of finding an explicit expression
for the eigenfunctions:
a systematic construction of the polynomial solutions to the
generalized Laplace equations has not been accomplished yet \cite{cal69}.
A special solution for $N=5$ has been presented by Gambardella
\cite{gam75}.
The formal approach using the operator algebra \cite{pol92,bri92}
yields expressions for wave functions that are simple only for
small number of particles in the system.
It has been indeed claimed \cite{bri92} that the expressions
quickly become cumbersome due to complicated sums
in the definition of operators.
Similar difficulties may happen if the wave functions for
particles with internal degrees of freedom
are constructed from the operators introduced by
Minahan and Polychronakos \cite{min93}.

The purpose of this Letter is to explicitly construct a
set of the {\it exact} eigenfunctions
for the $N$-{\it body} Calogero-Sutherland model where the particles
with SU($\nu$) internal degrees of freedom move in external harmonic
confining potential.
It turns out that the eigenfunctions obtained here cover the
{\it full energy spectrum} although they belong to a special set of
complete eigenfunctions.
To illustrate the approach,
we start with the construction of wave functions
for the single-component case.
We then turn to the case of particles
with the internal SU($\nu$) spin degrees of freedom,
and present the eigenfunctions
and the corresponding energy spectrum.

Let us introduce the Hamiltonian in units of
$\hbar^2/m$
for the inverse-square model
of spinless particles confined by a harmonic potential
$\frac{1}{2} m \omega_0^2 x^2$ \cite{cal69,sut71},
$$
H = - \frac{1}{2} \sum_{i=1}^{N} \frac{\partial^2}{\partial x_i^2}
    + \frac{m^2 \omega_0^2}{2 \hbar^2} \sum_{i=1}^{N} x_i^2
    + \sum_{j>i} \frac{\lambda(\lambda \mp 1)}{(x_j-x_i)^2}
,
\eqno(1)
$$
where $N$ is the total number of particles
and $\lambda \geq 0$ is a dimensionless constant of pair coupling.
The upper (lower) sign
in the interaction term holds for bosons (fermions).

We first recall a characteristic feature of the
wave function inherent in the $1/r^2$ models:
the ground-state wave function is of the Jastrow type,
expressed as a product of the Jastrow factor and Gaussian functions
\cite{sut71}.  It has been known
that the construction of this wave function exhibits
a remarkable similarity to that of Laughlin's wave function
for the fractional quantum Hall (FQH) states
with the filling $1/p$, ${\mit \Psi}_p =z^p
\times {\rm (Gaussians)}$, where
$z$ is a Vandermonde determinant \cite{laf83}.
In the FQH states, it is further known that the
wave functions for excited states
are obtained by multiplying an appropriate
polynomial to the ground-state wave function
\cite{laf83}.  One naturally expects that even for excited states,
the construction of eigenfunctions for the $1/r^2$ models
can be quite analogous to that for the FQH states.
It was actually demonstrated
that this is indeed the case for the systems
with periodic boundary conditions
\cite{sut71,hal88,sha88,hah92,wan92}.

Based on these observations,
we thus propose a following  set of the Jastrow-type
ansatz eigenfunctions for the system of bosons and fermions
described by the above Hamiltonian (1),
$$
{\mit \Psi} (x_1, x_2, \ldots, x_N) = |z|^\lambda z^\gamma
{\mit \Phi} (x_1, x_2, \ldots, x_N)
,
\eqno(2)
$$
where
$z=\prod_{j>i} (x_j-x_i)$ is the Vandermonde determinantal product,
$\gamma=0$ ($\gamma=1$) for bosons (fermions),
and $\mit \Phi$ is assumed to be a symmetric polynomial.
The second factor $z^\gamma$ is introduced to
generate the symmetry (antisymmetry) property of
the wave function for the boson (fermion) case.
Note that ${{\mit \Psi}}{(\lambda=0)}$ corresponds to a solution
of the noninteracting case.
Motivated by the analogy to the FQH states,
let us further assume that
${\mit \Phi}$ is a product of a polynomial $F$ and a function $G$,
namely ${\mit \Phi}=F G$.
The function $G$ is a product of Gaussians,
$$
G(x_1, \ldots, x_N)
= \prod_{i=1}^N {\rm exp}(- \frac{m \omega_0}{2 \hbar} x_i^2)
,
\eqno(3)
$$
which can naturally take into account the effects
of the harmonic confinement.
The polynomial $F$ introduced here,
$$
F(x_1, \ldots, x_N)
=\sum_{m_1+\ldots+m_N=I}
a_{m_1 \ldots m_N}
\prod_{i=1}^N H_{m_i} (\sqrt{\frac{m \omega_0}{\hbar}} x_i)
,
\eqno(4)
$$
is a linear combination of Hermite polynomials $H_{m_i}$
with unknown amplitudes $a_{m_1 \ldots m_N}$
which should satisfy the condition of symmetry,
$a_{m_1 \ldots m_k \ldots m_\ell \ldots m_N}
=a_{m_1 \ldots m_\ell \ldots m_k \ldots m_N}$.
Note that the excitations are labeled by the quantum
numbers in the polynomial $F$,
$m_i=0,1,2 \ldots$ ($i=1,2,\ldots,N$).
The sum in the definition of the polynomial $F$ is taken over
quantum numbers $m_i$ ($i=1,2,\dots,N$)
which satisfy $\sum_{i=1}^N m_i=I$,
where $I$ is a given non-negative integer.

We show that ansatz wave function (2) with (3) and (4) satisfies
the Schr\"{o}dinger equation
$H {\mit \Psi}= E {\mit \Psi}$
in the sector
$x_1 \leq x_2 \leq \ldots \leq x_N$.
The proof can be easily extended to
the whole configuration space.
The application of the Hamiltonian (1) on the ansatz eigenfunction (2)
results in the following expression,
$$
\frac{1}{{\mit \Psi}}
H {\mit \Psi}
= \frac{m \omega_0}{\hbar}
\left [
\frac {1}{2} (\lambda+\gamma) N(N-1) +  \frac{N}{2} + I
\right ]
- \frac{\lambda+\gamma}{F} \sum_{\ell>k} \frac{1}{x_\ell-x_k}
\left (
\frac{\partial F}{\partial x_\ell}-\frac{\partial F}{\partial x_k}
\right )
{}.
\eqno(5)
$$
We would like to eliminate the cross-term containing the
derivatives of the function $F$.
Such requirement leads to the additional condition imposed
on the amplitudes $a_{m_1 \ldots m_N}$ for any $k<\ell$,
which is satisfied when the polynomial $F$ contains two terms
with the amplitudes related by
$$
m_k a_{m_1 \ldots m_k \ldots m_\ell \ldots m_N}
=
(m_\ell + 1) a_{m_1 \ldots m_k-1 \ldots m_\ell+1 \ldots m_N}
\eqno(6)
$$
for any $m_k>0$ and $m_\ell<I$.
It is further required that the sum in the definition
of the polynomial $F$ is taken over {\it all} possible
combinations of integers $m_i$ satisfying
$\sum_{i=1}^N m_i=I$.
The conditional equation (6)
is solved by the factorization of amplitudes
$a_{m_1 \ldots m_N}$ (with $0!=1$),
$$
a_{m_1 \ldots m_N} = \prod_{i=1}^N \frac{1}{m_i !}
{}.
\eqno(7)
$$

Consequently, the
eigenfunction for the Calogero-Sutherland-type model (1)
of spinless particles moving in a harmonic potential
can be written as
$$
{\mit \Psi}
=
\prod_{j>i} [|x_j-x_i|^\lambda (x_j-x_i)^\gamma ]
\sum_{m_1+\ldots+m_N=I}
\prod_{i=1}^N \frac{1}{m_i !}
H_{m_i}(\sqrt{\frac{m \omega_0}{\hbar}} x_i)
{\rm exp}(-\frac{m \omega_0}{2 \hbar} x_i^2)
{}.
\eqno(8)
$$
The corresponding energy spectrum follows from (5),
$$
E(N;I)
= \frac {1}{2} \hbar \omega_0 [ (\lambda+\gamma) N(N-1) + N ]
+ \hbar \omega_0 I
{}.
\eqno(9)
$$
where a non-negative integer $I$ labels the excitation.
The ground-state wave function and the corresponding eigenenergy
(obtained for $I=0$) reproduce the known exact results \cite{sut71}.
Note that the case of noninteracting
particles (both fermions and bosons)
is achieved by $\lambda=0$.

The energy spectrum (9) is a sum of the $N$-particle correlated
ground-state energy and the spectrum of $N$ noninteracting fermions
or bosons in a harmonic well, as should be
expected \cite{bri92}. For example, the excitation energy
of the particle-hole type, which is labeled by a positive
integer $I$ with fixed $N$,
is independent of the interaction strength $\lambda$.
Although the set of eigenfunctions (8) provides us with a
special series of the eigenfunctions for $N$ particles,
it supplies {\it all} energy levels of the system as can be seen
when one compares the energy spectrum (9) with the results
deduced by other techniques \cite{bri92}.
Hence, the energy levels obtained from the wave function (8)
cover all the excitation energy.
It is, however, still open to count the degeneracy of each level within
the present approach. We note that one can employ other methods
\cite{bri92} for counting the degeneracy.

We now generalize the model for particles with SU($\nu$)
spin degrees of freedom.
The integrability and the ground-state wave function
(up to $N=6$ particles)
were shown for the model by Minahan and Polychronakos \cite{min93}.
Recently, we have constructed the ground-state wave
function for an arbitrary number of particles \cite{vac93}.
Using the approach outlined above,
let us now obtain a set of excited-state wave functions
and energy spectrum for particles with
SU($\nu$) internal degrees of freedom.
The SU($\nu$) integrable generalization of the Hamiltonian
in units of $\hbar^2/m$ takes the form \cite{hah92,pol92,hik93}
$$
H = - \frac{1}{2} \sum_{i=1}^{N} \frac{\partial^2}{\partial x_i^2}
    +  {m^2 \omega_0^2 \over 2 \hbar^2} \sum_{i=1}^{N} x_i^2
    + \sum_{j>i} \frac{\lambda(\lambda+P_{ij}^\sigma)}{(x_j-x_i)^2}
,
\eqno(10)$$
where the spin-exchange operator $P_{ij}^{\sigma}$
of particles $i$, $j$ has been introduced.
We propose the Jastrow-type ansatz eigenfunction,
namely the trial function reads
$$
{\mit \Psi}(x_1 \sigma_1, \ldots ,x_N \sigma_N) =
\left \{ \prod_{j>i} |x_j-x_i|^{\lambda+\gamma-1}
(x_j-x_i)^{\delta_{\sigma_j \sigma_i}-\gamma+1}
{\rm exp}
\left [
i\frac{\pi}{2}{\rm sgn}(\sigma_j-\sigma_i)
\right ]
\right \}
$$
$$
\times
\left [
\; \;
\sum_{m_1+\ldots+m_N=I}
\; \; \;
\prod_{i=1}^N
\frac{1}{m_i !}
H_{m_i}(\sqrt{\frac{m \omega_0}{\hbar}} x_i)
\; \;
\right ]
\prod_{i=1}^{N} {\rm exp}(-\frac{m \omega_0}{2\hbar}x_i^2)
,
\eqno(11)$$
where the indices $\sigma_i$ ($i=1,2,\ldots,N$) denote
the SU($\nu$) spin of each particle.
The exponent $\gamma$ is equal to 0 and 1 for bosons and fermions,
respectively.
The third factor is introduced to take into account for the
symmetry of SU($\nu$) degrees of freedom.

Here we make a brief comment on the wave function (11).
As in other models with $1/r^2$ interaction \cite{sut71,hah92},
for instance in the fermion case,
the eigenfunction ${\mit \Psi}{(\lambda=0,\gamma=1)}$
is a solution for noninteracting SU($\nu$) fermions
in a harmonic well.
So, the wave function (11) can be rewritten as
${\mit \Psi}=  \prod_{j>i} |x_j-x_i|^{\lambda}
{\mit \Psi}{(\lambda=0,\gamma=1)}$.
One can clearly see from this expression
the analogy to Jain's construction \cite{jai89}
of the wavefunction for hierarchical FQH states
in which the Jastrow factor is introduced to
non-interacting electrons of filled Landau levels
with the filling $\nu$.
Furthermore, following the proof for the
SU(2) case \cite{vac93},
it is easily checked that the eigenfunction $\mit \Psi$
is a product written as $ {\mit \Psi} = F {\mit \Psi}_G,$
where $F$ is defined by (4) and (7),
and ${\mit \Psi}_G$ is the ground-state wave function of
the Hamiltonian (10) for a given spin configuration.
This decomposition implies that we are now seeking for
the eigenfunctions whose symmetry is the same as
${\mit \Psi}_G $ since a completely symmetric
polynomial $F$ does not change the symmetry property.

It is now straightforward to show that
the substitution of the trial function (11)
into the Schr\"{o}dinger equation with the Hamiltonian (10)
yields the expression for the eigenenergy
$$
E(N_1,N_2, \ldots, N_\nu;I)
=
\frac{1}{2} \hbar \omega_0
\left [
\lambda
  N ( N - 1 )
+ \sum_{\alpha=1}^{\nu} N_\alpha^2
\right ]
+\hbar \omega_0 I
,
\eqno(12)
$$
where the configuration of spins is denoted by
the number of particles,
$N_1,N_2, \ldots, N_\nu$ ($\sum_{\alpha=1}^{\nu} N_\alpha=N$).
The trial function (11) is therefore an eigenfunction
of the Hamiltonian (10).
The correlations via $1/r^2$ interaction appear only in
the ground-state energy for a given spin configuration,
$E_G = \frac{1}{2} \hbar \omega_0
\left [\lambda  N ( N - 1 )+
\sum_{\alpha=1}^{\nu} N_\alpha^2 \right ]$.
The excitations do not include any effects of interactions provided
that the number of electrons $N_{\alpha}$ is kept fixed \cite{kaw93b}.
The level of the spin-independent particle-hole excitations
is determined by the quantum number
$I=\sum_{i=1}^N m_i$;
the lowest lying excitation is obtained for $I=1$,
the next one for $I=2$, etc.
The case of noninteracting fermions
is achieved for $\lambda=0$, whereas for bosons,
the case of $\lambda=0$ corresponds to particles
with an infinite hard core, as discussed
in \cite{sut71} and \cite{kaw91}.
Notice that $\lambda$ in our notation for the multicomponent
model (10) corresponds to $\lambda-1$ in the notation
of \cite{sut71} and \cite{kaw91}.

The spectrum of the Hamiltonian (1) or (10)
is the same as the spectrum of noninteracting particles
shifted only by the correlated ground-state energy
obtained previously \cite{vac93}.
Comparing to the spectrum suggested in
the hypothesis \cite{kaw93b},
one can see that the energies of all quantum levels are produced
directly from wave functions (11).
As is the case for the other $1/r^2$
models \cite{hal88,sut93,hal92}, the degeneracy of each energy level
is expected to  be given by that of
noninteracting SU($\nu$) particles, described by
independent sets of quantum numbers
$m_i$, $i=1,2,\ldots,N$.
In order to figure out the problem of degeneracy and
obtain the corresponding wave functions microscopically,
it is desirable to  analyze symmetry properties of the model
in detail \cite{bri92,min93}.
In particular, we think that Yangian symmetry
discussed in \cite{hal92} may play a key role for the classification
of the complete spectrum of the present SU($\nu$) model.

In conclusion, we have presented a set of the excited-state wave functions
for SU($\nu$) generalization of the quantum Calogero-Sutherland
Hamiltonian with harmonic confinement.
Although the present wave functions belong to a special subset of
complete eigenfunctions, it is remarkable that the constructed set of
eigenfunctions provides us exactly with all the energy levels for
the excited states.

Fruitful discussions with F. D. M. Haldane, Y. Kuramoto, B. S. Shastry,
and critical reading of the manuscript by W. Brenig
are gratefully acknowledged.
This work was partly supported by
a Grant-in-Aid from the Ministry of Education, Science and Culture.
K. V. acknowledges support from the Japanese
Government (Monbusho) Scholarship Program, and N. K. acknowledges
support from the Monbusho International Scientific Research Program.



\begin{thebibliography}{99}
\bibitem{fra93}
The spectrum for the SU($\nu$) spin chain has also been studied
numerically by H. Frahm, J. Phys. A {\bf 26}, L473 (1993) under a
specific condition for the positions of particles.
\end{thebibliography}
\end{document}